\documentclass[prb,twocolumn,showpacs,amsmath,amssymb,superscriptaddress,preprintnumbers]{revtex4}
\usepackage{graphicx}% Include figure files
\usepackage{dcolumn} % Align table columns on decimal point
\usepackage{bm}
\usepackage{epsfig}

\begin{document}

\title{Optical properties and magnetic field-induced phase transitions
in the ferroelectric state of Ni$_3$V$_2$O$_8$}

\author{{R. C. Rai}}
\affiliation{Department of Chemistry,
University of Tennessee, Knoxville, TN 37996}
\author{{J. Cao}}
\affiliation{Department of Chemistry, University of Tennessee,
Knoxville, TN 37996}
\author{{S. Brown}}
\affiliation{Department of Chemistry, University of Tennessee,
Knoxville, TN 37996}
\author{J. L. Musfeldt}
\email{musfeldt@utk.edu}\affiliation{Department of Chemistry, University of Tennessee,
Knoxville, TN 37996}
\author{D. Kasinathan}
\affiliation{Department of Physics, University of California Davis,
Davis, CA 95616}
\author{{D. J. Singh}}
\affiliation{Materials Science and Technology Division, Oak Ridge
National Laboratory, Oak Ridge,TN 37831-6032}
\author{G. Lawes}
\affiliation{Department of Physics, Wayne State University, Detroit,
MI 48201}
\author{N. Rogado}
\affiliation{DuPont Central Research and Development, Experimental
Station, Wilmington, DE 19880-0328}
\author{R. J. Cava}
\affiliation{Department of Chemistry and Princeton Materials
Institute, Princeton University, Princeton, NJ 08544}
\author{{X. Wei}}
\affiliation{National High Magnetic Field Laboratory, Florida State
University, Tallahassee, Florida 32310}


\begin{abstract}
We use a combination of optical spectra, first principles
calculations, and energy dependent magneto-optical measurements to
elucidate the electronic structure and to study the phase diagram of
Ni$_3$V$_2$O$_8$. We find a remarkable interplay of magnetic field
and optical properties that reveals additional high magnetic field
phases and an unexpected electronic structure which we associate
with the strong magneto-dielectric couplings in this material over a
wide energy range. Specifically, we observed several prominent
magneto-dielectric effects that derive from changes in crystal field
environment around Ni spine and cross-tie centers. This effect is
consistent with a field-induced modification of local structure.
Symmetry-breaking effects are also evident with temperature. We find
Ni$_3$V$_2$O$_8$ to be an intermediate gap, local moment band
insulator. This electronic structure is particularly favorable for
magneto-dielectric couplings, because the material is not subject to
the spin charge separation characteristic of strongly correlated
large gap Mott insulators, while at the same time remaining a
magnetic insulator independent of the particular spin order and
temperature.
\end{abstract}

\maketitle

\section{INTRODUCTION}

Ni$_3$V$_2$O$_8$ is a particularly interesting magnetic material,
\cite{sauerbrei73,rogado02,lawes05,lawes04,kenzelmann06} both
because of its unusual structure, which provides an example of a
spin 1 system on a Kagom$\acute{e}$ staircase lattice, and because
of the rich variety of magnetic and structural phases that are
stabilized under different conditions. One especially interesting
feature is the occurrence of a magnetic, ferroelectric phase as a
function of temperature and magnetic field. More generally, coupled
magnetic and electric degrees of freedom, flexible lattices, and
magnetic frustration in multiferroics can result in cascades of
coupled magnetic and dielectric transitions.
\cite{cheong04,cruz05,lorenz04,tgoto04,hur04,lawes03,rogado05,saito05}
The recent report \cite{hur04} of colossal low frequency (1 kHz)
magneto-dielectric effects in inhomogeneously mixed-valent
DyMn$_2$O$_5$ is especially important, because it illustrates that
sizable dielectric contrast can be achieved by physical tuning
through an unusual commensurate-incommensurate magnetic transition
and is facilitated by a soft lattice. The 300 K low-frequency
magneto-dielectric effect in mixed-valent LuFe$_2$O$_4$ has also
attracted attention due to the very low magnetic fields needed to
achieve dielectric contrast.\cite{subra06} Ni$_3$V$_2$O$_8$ is
another system where the temperature and field dependence of the
spontaneous polarization shows a strong coupling between magnetic
and ferroelectric order. \cite{lawes05,lawes04,kenzelmann06} This
coexistence is unusual and appears only when certain symmetry
conditions are fulfilled. \cite{lawes05} That the effect can be
controlled with an external magnetic field makes it attractive for
device applications. Based upon our previous work with
inhomogeneously mixed-valent K$_2$V$_3$O$_8$,\cite{rai06} the
significant coupling between spin, lattice, and charge degrees of
freedom make Ni$_3$V$_2$O$_8$ an excellent candidate for discovery
of higher energy magneto-dielectric effects.

\begin{figure}[b]
\centerline{
\includegraphics[width=3.6in]{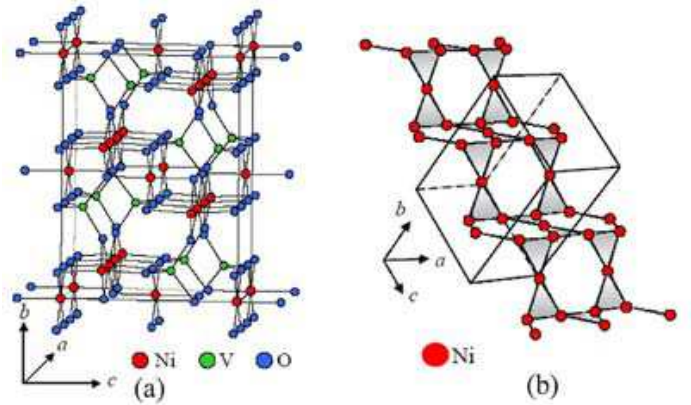}}
\vskip -0.2cm \caption{(Color online) (a) Crystal structure of
Ni$_3$V$_2$O$_8$. (b) View of Kagom$\acute{e}$ staircase showing
only the Ni atoms.}\label{struct}
\end{figure}

\begin{figure}[h]
\centerline{
\includegraphics[width=3.0in]{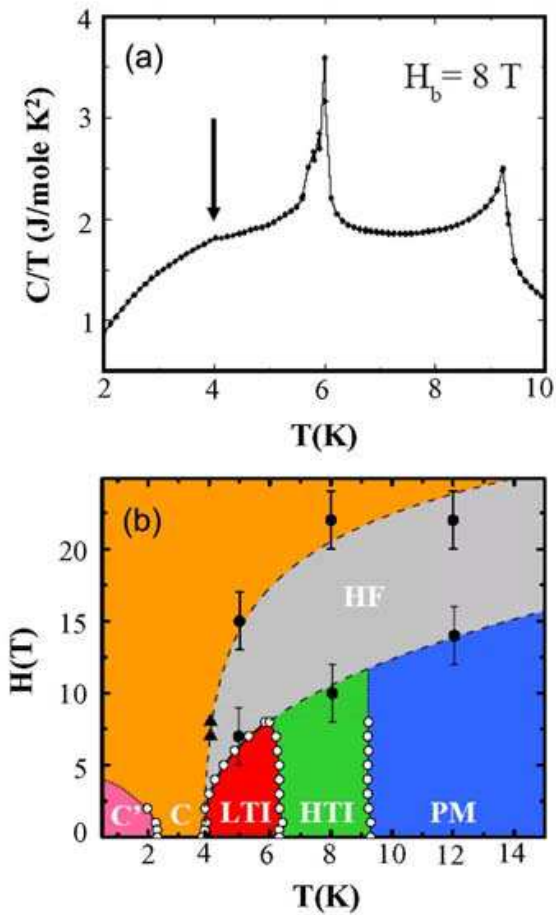}}
\vskip -0.2cm \caption{(Color online) (a) Specific heat of
Ni$_3$V$_2$O$_8$ as a function of temperature at $H$ = 8 T. The
arrow indicates a proposed phase transition temperature. (b) $H$-$T$
phase diagram of Ni$_3$V$_2$O$_8$ for $H$$\parallel$$b$ with
additional data points extracted from specific heat
(solid triangles) and  magneto-optical measurements (solid
circles). Solid lines are discussed in the text and the dashed line
is a guide to the eye. A new high field (HF) phase is also
indicated.}\label{PhaseDia}
\end{figure}

Figure \ref{struct}(a) shows the orthorhombic (Cmca) crystal
structure of Ni$_3$V$_2$O$_8$. It consists of Kagom$\acute{e}$
layers of edge sharing NiO$_6$ octahedra separated by nonmagnetic
VO$_4$ tetrahedra. Ni$_3$V$_2$O$_8$ is considered to be a
Kagom$\acute{e}$ staircase compound due to buckling of the lattice
perpendicular to the $a$ axis. There are two distinct types of
Ni$^{2+}$ ($S$ = 1) centers, which we refer to as ``spine" and
``cross-tie" sites. The ``spine" sites run along the $a$ axis. A
view of the Kagom$\acute{e}$ staircase showing only the Ni atoms is
displayed in Fig. \ref{struct}(b). Note that the spine and cross-tie
sites have very different local symmetries.  The spin ordering
arrangements in Ni$_3$V$_2$O$_8$ have been extensively investigated
by neutron scattering and derive from various local and long range
exchange, spin anisotropy, Dzyaloshinskii-Moriya, dipolar, and
frustration effects. \cite{kenzelmann06} The $H$-$T$ phase diagram
(for $H$$\parallel$$b$) is shown in Fig. \ref{PhaseDia}(b). Two
incommensurate phases are observed below the paramagnetic phase
(PM). HTI is the (longitudinal) high temperature incommensurate
phase, and LTI is the (spiral) low temperature incommensurate phase.
The latter displays ferroelectricity. Commensurate canted
antiferromagnetism is observed in the C phase.

Here we use a combination of optical spectroscopy, first principles
calculations, and energy dependent magneto-optical measurements to
elucidate the electronic structure and to study the phase diagram of
Ni$_3$V$_2$O$_8$. We find a remarkable interplay of magnetic field
and optical properties that reveals additional high magnetic field
phases and an unexpected electronic structure which we associate
with the strong magneto-dielectric couplings in this material over a
wide energy range. Specifically, we observed several prominent
magneto-dielectric effects that derive from changes in crystal field
environment around Ni spine and cross-tie centers. This effect is
consistent with a field-induced modification of local structure.
Symmetry-breaking effects are also evident with temperature. Even
though both Ni$_3$V$_2$O$_8$ and the prototypical Mott insulator NiO
are based on Ni$^{2+}$ ions octahedrally coordinated by O with
similar bond lengths, we show that the basic electronic structures
of these two materials are very different. NiO has a large band gap
and local moments that derive from strong Coulomb interactions and
separation of spin and charge degrees of freedom. This separation is
theoretically interesting but disadvantageous if one wants to
promote coupling effects. The Slater insulator is at the opposite
extreme. Here, a small band gap is present in the ground state, but
it is a direct result of a specific spin ordering. In contrast, we
find Ni$_3$V$_2$O$_8$ to be an intermediate gap, local moment band
insulator. This electronic structure is particularly favorable for
magneto-dielectric couplings, because the material is not subject to
the spin charge separation characteristic of strongly correlated
large gap Mott insulators, while at the same time remaining a
magnetic insulator independent of the particular spin order and
temperature.

\section{Methods}

\subsection{Crystal Growth}

The Ni$_3$V$_2$O$_8$ single crystal samples were prepared from a
BaO-V$_2$O$_5$ flux.  The crystals used for these measurements were
grown as platelets, with their largest faces (a few square
millimeters) perpendicular to the crystallographic $b$ axis.

\subsection{Spectroscopic Investigations}

Near normal $ac$-plane  reflectance of Ni$_3$V$_2$O$_8$ was measured
over a wide energy range (3.7 meV - 6.5 eV) using several different
spectrometers including a Bruker 113 V Fourier transform infrared
spectrometer, a Bruker Equinox 55 Fourier transform infrared
spectrometer equipped with an infrared microscope, and a Perkin
Elmer Lambda 900 grating spectrometer, as described previously.
\cite{zhu02} The spectral resolution was 2 cm$^{-1}$ in the far and
middle-infrared and 2 nm in the near-infrared, visible, and
near-ultraviolet. Optical conductivity was calculated by a
Kramers-Kronig analysis of the measured reflectance.\cite{wooten72}
An open flow cryostat  provided
temperature control.

The magneto-optical properties of Ni$_3$V$_2$O$_8$ were investigated
between 0.75 and 4.1 eV using a 3/4 m grating spectrometer equipped
with InGaAs and CCD detectors and a 33 T resistive magnet at the
National High Magnetic Field Laboratory (NHMFL), in Tallahassee, FL.
150, 300, and 600 line/mm gratings were used, as appropriate.
Experiments were performed with polarized light ($E$$\parallel$$a$
and $E$$\parallel$$c$) in the temperature range between 4.2 and 15 K
for applied magnetic fields up to 30 T ($H$$\parallel$$b$). With
light shining on the sample, we estimate that the base temperature
was $\sim$5.0 K. The field-induced changes in the measured
reflectance were studied by taking the ratio of the reflectance at
each field with the reflectance at zero field, i.e., [R($H$)/R($H$ =
0 T)]. This normalized response is a sensitive way to view the
field-induced optical changes.\cite{cao06} Since
$\epsilon(E)=\epsilon_1(E) + i\epsilon_2(E)
=\epsilon_1(E) + {{4{\pi}i}\over {\omega}}\sigma_1(E)$, it
is clear that the field-induced changes in reflectance translate
into finite energy magneto-dielectric effects. To obtain the 30 T
optical conductivity ($\sigma_1$) and dielectric response
($\epsilon_1$), we renormalized the zero-field absolute reflectance
with the high-field reflectance ratios, and recalculated
$\sigma$$_1$ and $\epsilon_1$ using Kramers-Kronig
techniques.\cite{wooten72,rai06}

\subsection{Electronic Structure Calculations}

The calculations were done with the experimental crystal structure
\cite{sauerbrei73} using the general potential linearized augmented
planewave (LAPW) method, with local orbitals, \cite{lapw,lo} as
implemented in the WIEN2K program. \cite{wien} The augmented
planewave plus local orbital extension was used for the O $s$ and
$p$ states, the metal $d$ states and the semi-core levels.
\cite{apw} The valence states were treated in a scalar relativistic
approximation, while the core states were treated relativistically.
The calculations were done self-consistently using well converged
basis sets and zone samplings based on 144 {\bf k}-points in the
irreducible wedge. A more dense mesh of 1200 {\bf k}-points was used
for the calculations of the optical conductivity. The LAPW sphere
radii were 1.8, 1.7 and 1.5 $a_0$ for Ni, V and O, respectively.
Calculations were done in the local spin density approximation
(LSDA) and the LDA+U method. For the LDA+U calculations, we used
values of $U$ ranging from 5 eV to 7 eV on the Ni sites. The results
shown are for $U$ = 5 eV. \cite{u-note}

\section{Results and Discussion}

\subsection{Optical and Electronic Properties of Ni$_3$V$_2$O$_8$}

\begin{figure}[t]
\centerline{
\includegraphics[width=3.2in]{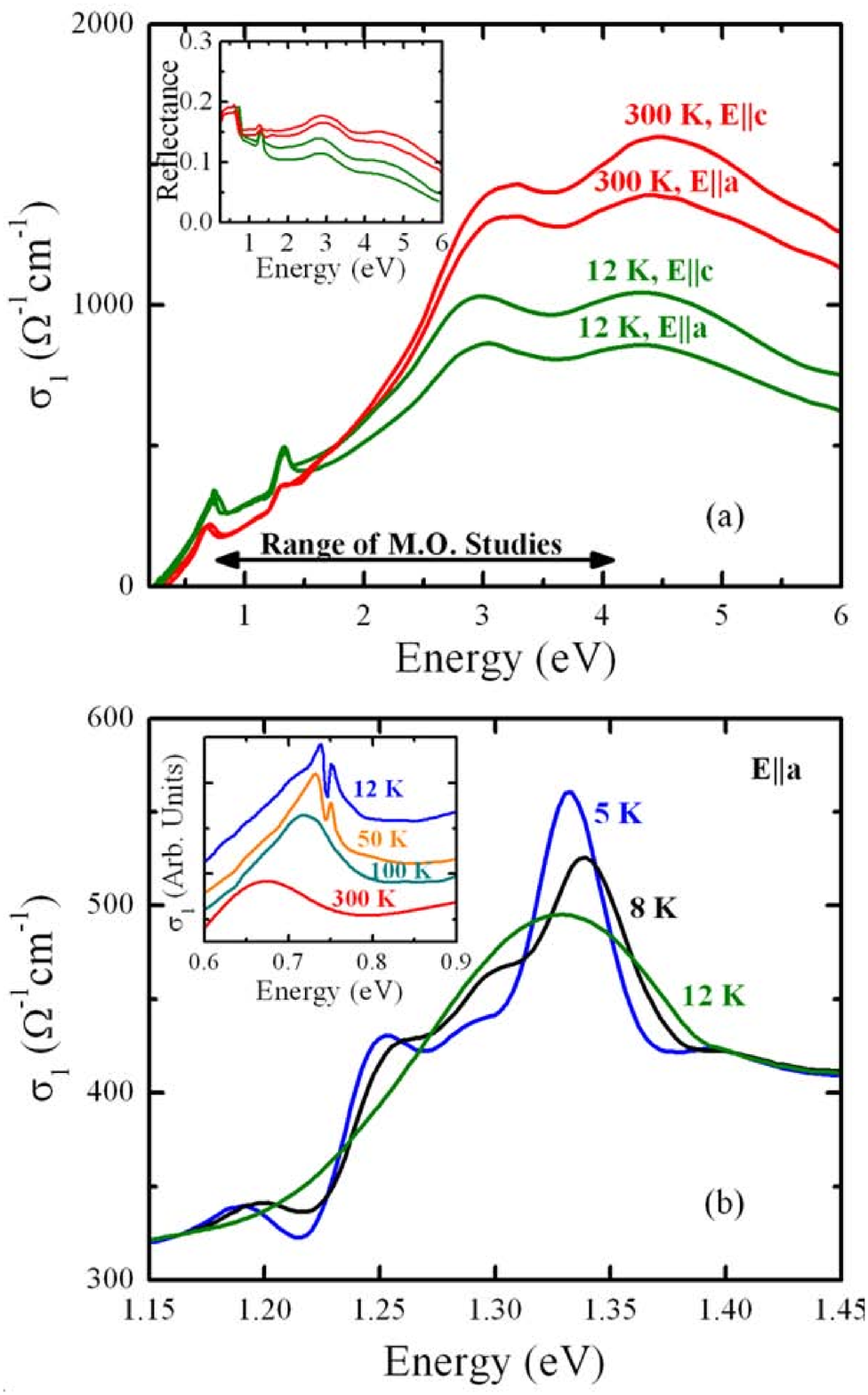}}
\vskip -0.2cm \caption{(Color online) (a) Polarized optical
conductivity of Ni$_3$V$_2$O$_8$ at 300 and 12 K, extracted from
reflectance measurements (inset) by a Kramers-Kronig analysis. The
energy range of our magneto-optical measurements is indicated by the
arrow (0.75 - 4.1 eV). (b) A close-up view of the $E$$\parallel$$a$
optical conductivity at 12, 8, and 5 K, corresponding to Ni (spine)
$d$ to $d$ on-site excitations. Inset shows a close-up view of Ni
(cross-tie) $d$ to $d$ on-site excitations at $T$ = 300, 100, 50,
and 12 K, respectively. These curves are offset for clarity.
}\label{OpConducH0}
\end{figure}

Figure \ref{OpConducH0}(a) displays the polarized optical
conductivity of Ni$_3$V$_2$O$_8$ in the PM phase at 300 and 12 K.
The spectra show directionally dependent vibrational and electronic
excitations with an optical energy gap of $\sim$0.35 eV. Based on
our electronic structure calculations (discussed in detail below)
and comparison with chemically similar Ni-containing
compounds,\cite{sawatzky,schnack06,ido91,juza66} we assign the
excitations centered near $\sim$0.75  and $\sim$1.35 eV to Ni $d$ to
$d$ on-site excitations in the minority spin channel on cross-tie
and spine sites, respectively. These excitations are optically
allowed due to the modest hybridization between the Ni $d$ and O $p$
states. We assign the $\sim$3.0 eV feature in the optical
conductivity spectrum to a combination of O 2$p$ to Ni 3$d$ and O
2$p$ to V 3$d$ charge transfer excitations. The $\sim$4.4 eV feature
derives from O 2$p$ to V 3$d$ charge transfer excitations.

Transition metal $d$ to $d$ on-site excitations are sensitive
indicators of the local crystal field environment. Figure
\ref{OpConducH0}(b) shows a close-up view of the Ni (spine) $d$ to
$d$ on-site excitations in the $E$$\parallel$$a$ optical
conductivity at 12, 8, and 5 K. This structure is broad and
featureless in the PM phase (12 K), but it splits into at least four
different components at 8 K (HTI phase). The splitting becomes more
pronounced at 5 K (LTI phase). This splitting is due to a local
structure distortion around the Ni center with decreasing
temperature and demonstrates that the NiO$_6$ environment is
different in the PM, HTI, and LTI phases. Although it is at the
limit of our detector range, the trailing edge of the $\sim$0.75
feature also displays splitting below 12 K (not shown).\cite{note1}
Interestingly, the Ni (cross-tie) $d$ to $d$ on-site excitation also
shows a notch-like structure that develops between 100 and 50 K
(inset of Fig. \ref{OpConducH0}(b)), indicating that a weak
structural distortion in the cross-tie direction precedes the
cascade of low temperature magnetic phases. Anomalous intermediate
temperature phonon shifts have also been reported in the
RMn$_2$O$_5$ (R = Bi, Eu, Dy) family of frustrated
multiferroics.\cite{flores06}

\subsection{Electronic Structure Calculations of Ni$_3$V$_2$O$_8$}

As mentioned above, there are two nonequivalent Ni sites in the unit
cell, a spine site (4 atoms per cell, denoted Ni2 in the following)
and a cross-link site (2 atoms per cell, denoted Ni1). As mentioned,
Ni$_3$V$_2$O$_8$ is a local moment magnet with a complex magnetic
phase diagram. To assign the peaks in the optical spectrum, we did
calculations for two magnetic structures: (1) all spins on each Ni
sublattice aligned ferromagnetically, but with the two sublattices
opposite to each other (denoted FiM, in the following) and (2) a
ferromagnetic ordering (FM). In the LSDA, the FiM ordering had lower
energy by 7 meV per formula unit (note that there are two formula
units per cell). This shows an antiferromagnetic interaction between
the two sublattices.

\begin{figure}[h]
\epsfig{width=0.85\columnwidth,file=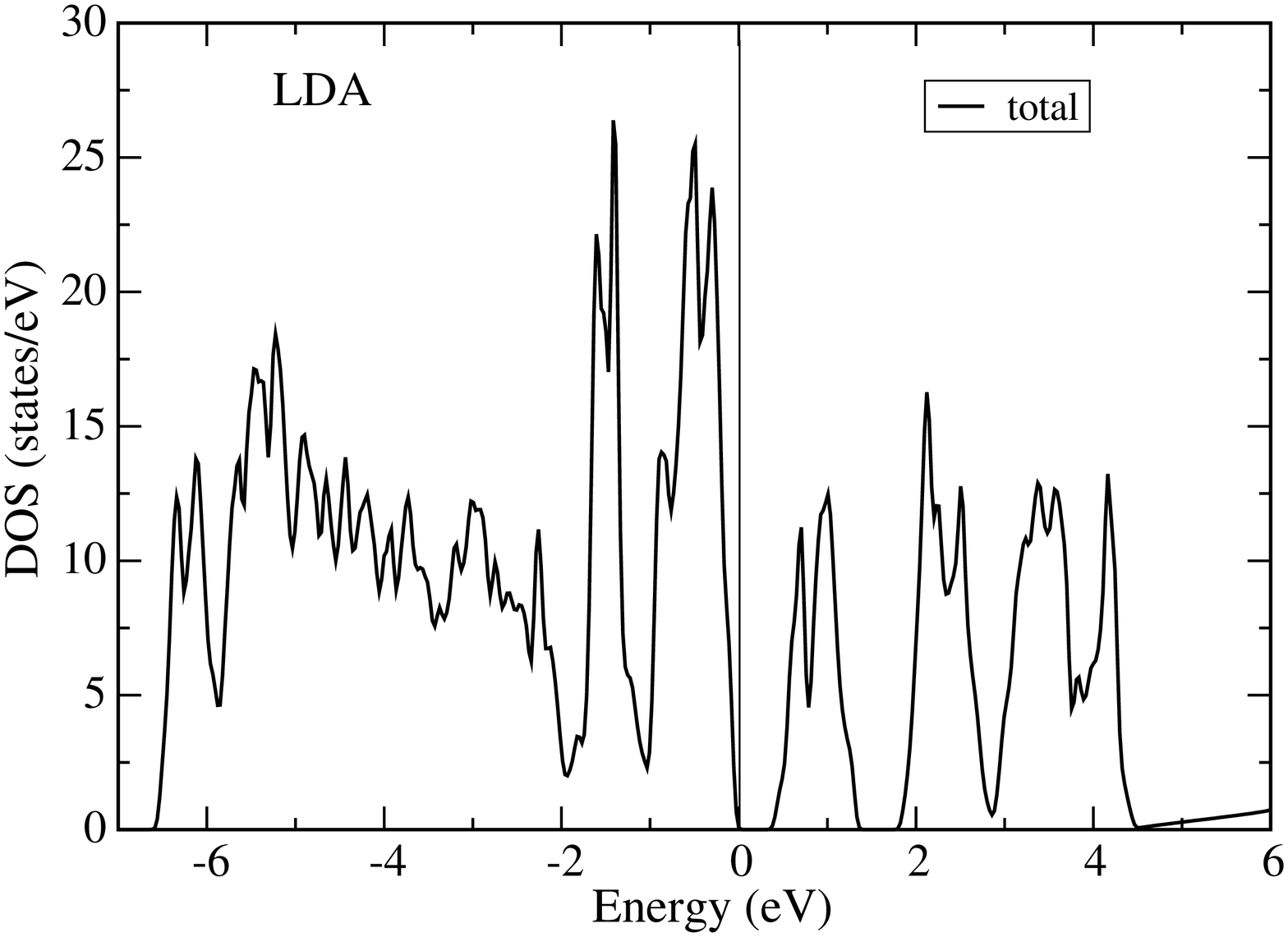} \vspace{.2cm}
\epsfig{width=0.85\columnwidth,file=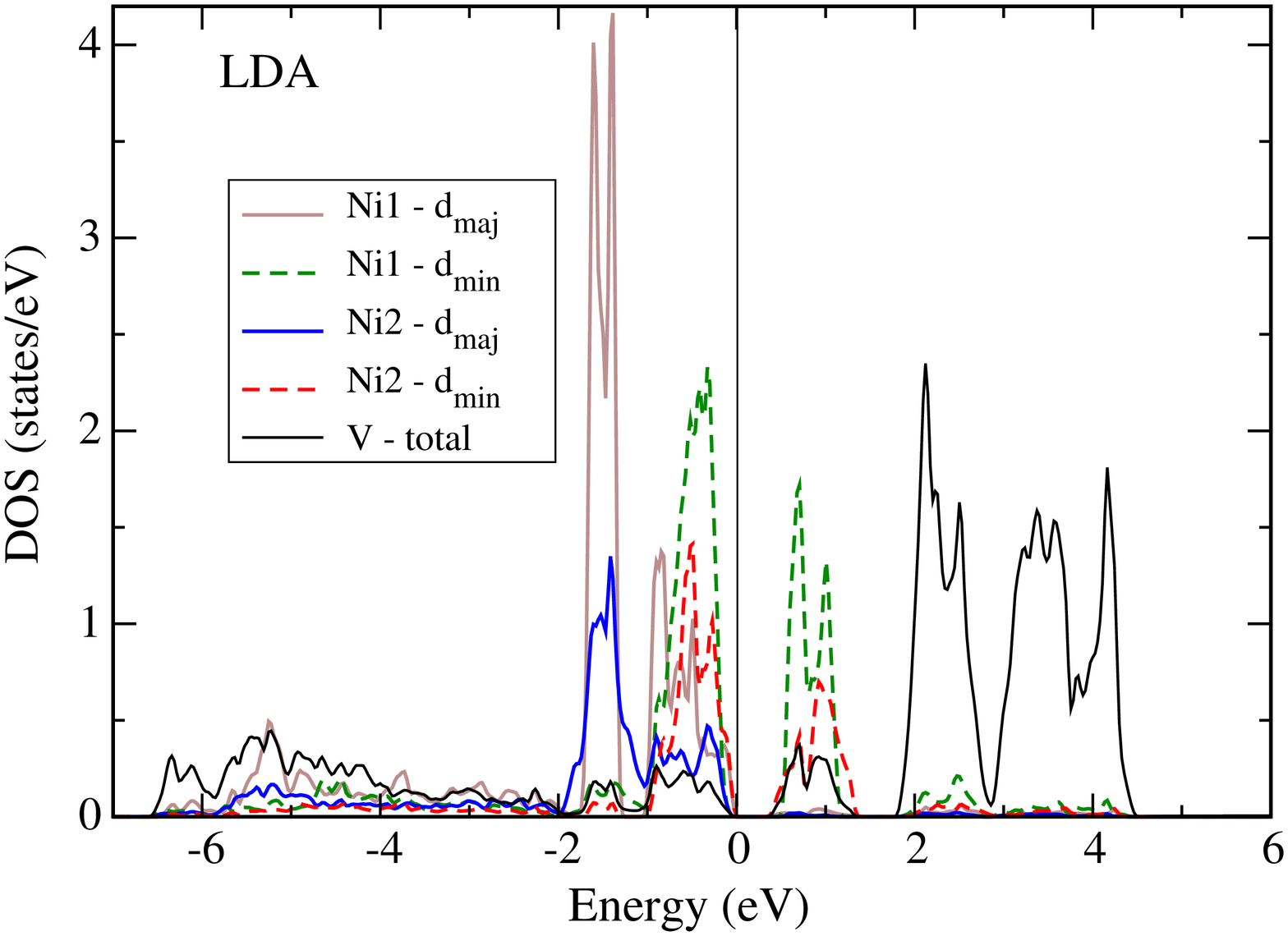}
\vspace{0.2cm} \caption{(color online) Total (top) and projected DOS
(bottom) as obtained with the LSDA for the FiM ordering (see text).
The total DOS is on a per formula unit spin basis. The projections
are onto the LAPW spheres, and are given per atom. } \label{dos-lda}
\end{figure}

\begin{figure}[b]
\epsfig{width=0.85\columnwidth,file=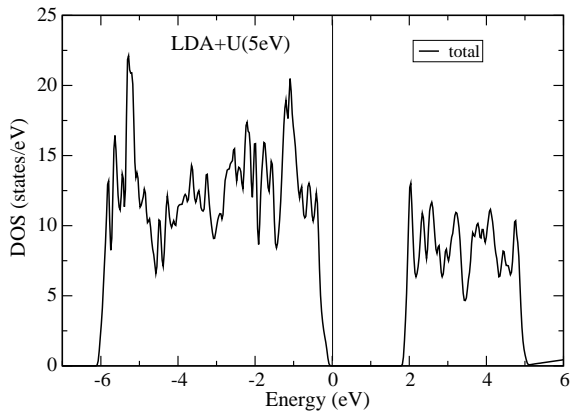}
\vspace{0.2cm}
\epsfig{width=0.85\columnwidth,file=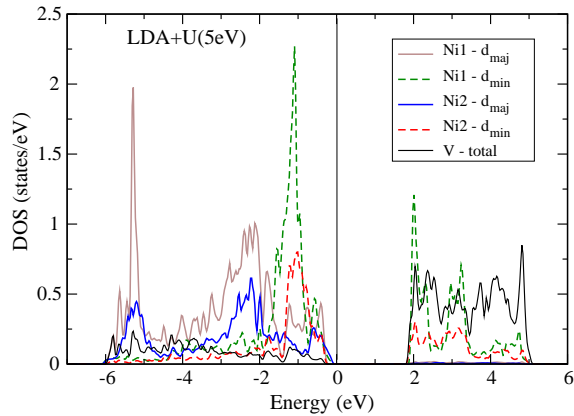}
\vspace{0.2cm} \caption{(color online) Total (top) and projected DOS
(bottom), as in Fig. \ref{dos-lda}, but with the LDA+U method, $U$ =
5 eV. } \label{dos-ldau}
\end{figure}

The calculated total and projected densities of states (DOS) for the
FiM ordering are shown in Fig. \ref{dos-lda} and Fig.
\ref{dos-ldau}, within the LSDA and the LDA+U (U = 5.0 eV)
approximation, respectively. In addition to the U = 5 eV DOS shown,
we did calculations for U = 6 eV and U = 7 eV (see below).
Insulating behavior is found in both approximations. The gaps are
$E_g({\rm LSDA})$ = 0.30 eV and $E_g({\rm LDA+U})$ = 1.90 eV. There
are only relatively small changes in the O $2p$ bands between these
two approximations. The O 2$p$ valence bands extend from $\sim$ -6.5
eV to the valence band edge in the LSDA, and  from $\sim$ -6.0 eV to
the band edge in the LDA+U approximation. Wang and co-workers
\cite{wang} investigated Ni$_3$V$_2$O$_8$ as a potential
photocatalyst for water splitting using optical spectroscopy from
300 nm to 850 nm, and density functional calculations. However, they
did not include spin-polarization in their band structure
calculations, and this is needed to obtain the exchange splitting of
the Ni $d$ states. As a result, they obtained a metallic electronic
structure with a high density of Ni $d$ states at the Fermi energy
in disagreement with experiment.

The compound is described as Ni$^{2+}_3$V$^{5+}_2$O$^{2-}_8$. As
shown in the projected DOS, we find the V $d$ bands well above
valence bands in both the LSDA and LDA+U calculations (from $\sim$ 2
eV to $\sim$ 5 eV, relative to the valence band edge), in agreement
with this ionic model. In all calculations, we find integer spin
magnetizations of 2 $\mu_B$ per Ni ion, as expected for Ni$^{2+}$.
In the crystal field, the main Ni $d$ splitting is due to the local
coordination, which gives a lower 3-fold degenerate $t_{2g}$ and a
higher two-fold degenerate $e_g$ manifold per spin. The Ni $d$ bands
are already very narrow in the LSDA. This is the reason for the
clean gap. In the FM case, the band gap $E_g$ is reduced by only
0.11 eV relative to the FiM case. Presumably, antiferromagnetism
within the individual sublattices would narrow the bands further and
increase $E_g$ relative to the FiM ordering. Thus, already at the
LSDA level, Ni$_3$V$_2$O$_8$ is a local moment insulator. The
majority spin Ni $d$ levels are centered at -1.7 eV ($t_{2g}$) and
-0.6 eV ($e_g$), for both Ni sites. The corresponding minority spin
levels are centered at -0.4 eV ($t_{2g}$) and 0.8 eV ($e_g$).

\begin{figure}
\epsfig{width=0.85\columnwidth,file=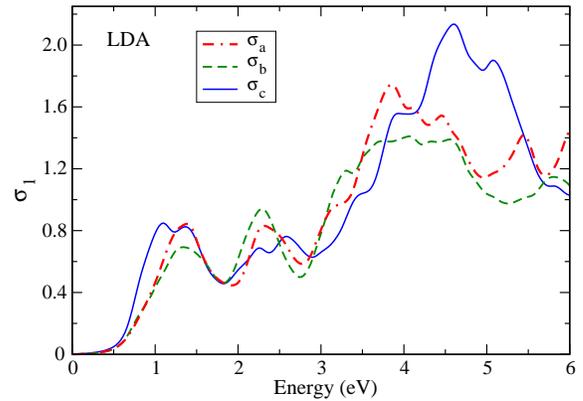}
\vspace{0.2cm}
\epsfig{width=0.85\columnwidth,file=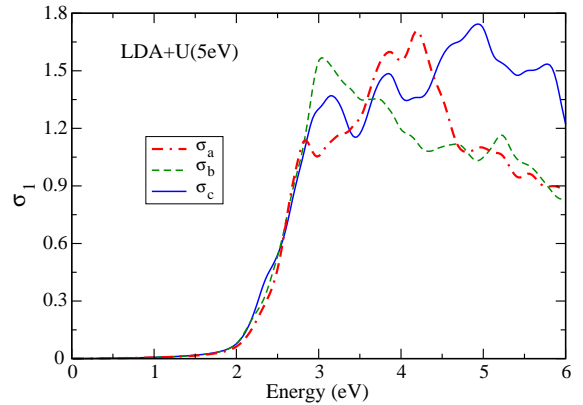}
\vspace{0.2cm} \caption{Real part of the optical conductivity within
the LSDA (top) and LDA+U, $U$ = 5 eV (bottom), for Ni$_3$V$_2$O$_8$
with the FiM ordering. A 0.1 eV broadening was applied. Two
definitions of the crystallographic axes are in literature. We use
the one where $a$ = 5.936 \AA, $b$ = 11.420 \AA, and $c$ = 8.240
\AA.\cite{sauerbrei73}} \label{sigma}
\end{figure}

The main difference between the LSDA and LDA+U electronic structure
is in the position of the Ni $d$ bands. The majority spin Ni
manifolds are shifted deep into the valence band. These are centered
at $\sim$ -5.5 eV ($t_{2g}$) and $\sim$ -2.5 eV ($e_g$). The
minority spin states are split apart by the Coulomb interaction and
are now centered at $\sim$ -1 eV ($t_{2g}$) and $\sim$ 2.5 eV
($e_g$). For this value of $U$ the conduction band edge onset
derives from both V $d$ and Ni $d$ states. Calculations with $U$ = 6
eV and $U$ = 7 eV increase the differences from the LSDA, pushing
the Ni $d$ conduction band onset above the V $d$ onset and driving
the occupied Ni $d$ manifolds deeper into the O $p$ bands.

The calculated LSDA and LDA+U optical spectra for FiM ordering are
shown in Fig. \ref{sigma}. Besides the larger LDA+U band gap, the
shapes of the spectra are very different. In the LSDA the onset has
both Ni $d$ - Ni $d$ and charge transfer character. The first
structure just above 1 eV is due to the structure in the Ni $d$ to
Ni $d$ excitations. It contains two peaks. The Ni1 (cross-link) site
has a lower primary crystal field splitting than the Ni2 (spine)
site, as may be seen from the projected DOS. Therefore, the lower
energy peak comes more from the Ni1 and the higher peak from the
Ni2. The second main structure above 2 eV, which is weaker, is of
mainly charge transfer character involving Ni and V, and the onset
starting at 3 eV is mainly of charge transfer character into the V
$d$ bands. In the LDA+U case the spectrum starts at $\sim$ 2 eV and
is dominated throughout by charge transfer excitations.

The LSDA spectrum is clearly in better accord with the experimental
spectrum than the LDA+U spectrum, both in terms of the gap and in
terms of the peak structure. The main differences between LSDA and
experiment are in the region of the second peak (above 2 eV) and in
the onset of the O $2p$ to V $d$ charge transfer excitations.
Considering that V is in a $d^0$ configuration, the underestimate of
this gap is not surprising and presumably reflects just the generic
LDA band gap underestimation, which is seen in other $d^0$ oxides.
The relatively good agreement of the LSDA Ni $d$ position and the
fundamental gap with experiment is, however, unexpected and deserves
comment.

NiO, which like Ni$_3$V$_2$O$_8$, has Ni$^{2+}$, six-fold
coordinated by O, is a prototypical Mott insulator. LSDA
calculations for antiferromagnetic NiO show a small band gap and
features in the valence band dispersions in accord with experiment.
\cite{terakura} However, the gap obtained is much smaller than the
experimental gap of $\sim$ 4 eV, and furthermore it is dependent on
the specific magnetic ordering. Furthermore, the LSDA gap of NiO is
of incorrect $d$-$d$ character rather than the charge transfer
character observed in experiment. \cite{sawatzky,merlin} This shows
an essential role for beyond LSDA Coulomb correlations in describing
the electronic structure of NiO. Various methods, including the
LDA+U approximation, for incorporating these have been developed and
tested using NiO.
\cite{norman,anisimov-1,anisimov-2,svanne,arai,kod,aryasetiawan,massidda,li}
As mentioned, NiO and Ni$_3$V$_2$O$_8$ both have Ni$^{2+}$ in
approximately octahedral O cages. In NiO, the Ni band width is
larger than in Ni$_3$V$_2$O$_8$, which explains why the LSDA is able
to produce local moment insulating behavior in this compound but not
in NiO. The narrower bands in Ni$_3$V$_2$O$_8$ are related to the
bonding topology. In NiO, the O are six fold coordinated with
90$^\circ$ and straight 180$^\circ$ bonds. Ni$_3$V$_2$O$_8$ has
lower O coordination and bent bonds. In metals, the criterion for
dividing strongly correlated materials from band metals involves
$U_s/W$, where $W$ is the band width and $U_s$ is the screened
Coulomb potential, where the ``s" is simply to distinguish this
parameter from the $U$ used in the LDA+U approximation. In general,
Mott insulators derived from metals with $U_s/W$ cannot be described
in the LSDA and require the use of the LDA+U approximation or more
sophisticated approaches, while materials with weaker correlations
are often better described in LSDA than in LDA+U calculation. Since
the band width of Ni$_3$V$_2$O$_8$ is not larger than that of NiO,
the implication of our results is that the effective screened
Coulomb interaction $U_s$ on at least one of the Ni sites is
sufficiently lower than in NiO to cross-over from strongly
correlated Mott insulator to band insulator. We speculate that this
could be due to the lower cation coordination of O$^{2-}$ in
Ni$_3$V$_2$O$_8$. Specifically, in NiO, six Ni atoms compete for
bonding with each O (two per O $p$ orbital). The average Ni-O bond
length in Ni$_3$V$_2$O$_8$ is $\sim$ 0.03 \AA$~$ shorter than in
NiO, consistent with this from the bond valence point of view.
\cite{brown} This is not the case in Ni$_3$V$_2$O$_8$, which may
enable the O cage to more effectively screen the bare $U$. As
mentioned, we find Ni$_3$V$_2$O$_8$ to be a local moment insulator
with weak sensitivity of the band structure to the magnetic order at
the LDA level.

\begin{figure}[h]
\epsfig{width=0.85\columnwidth,file=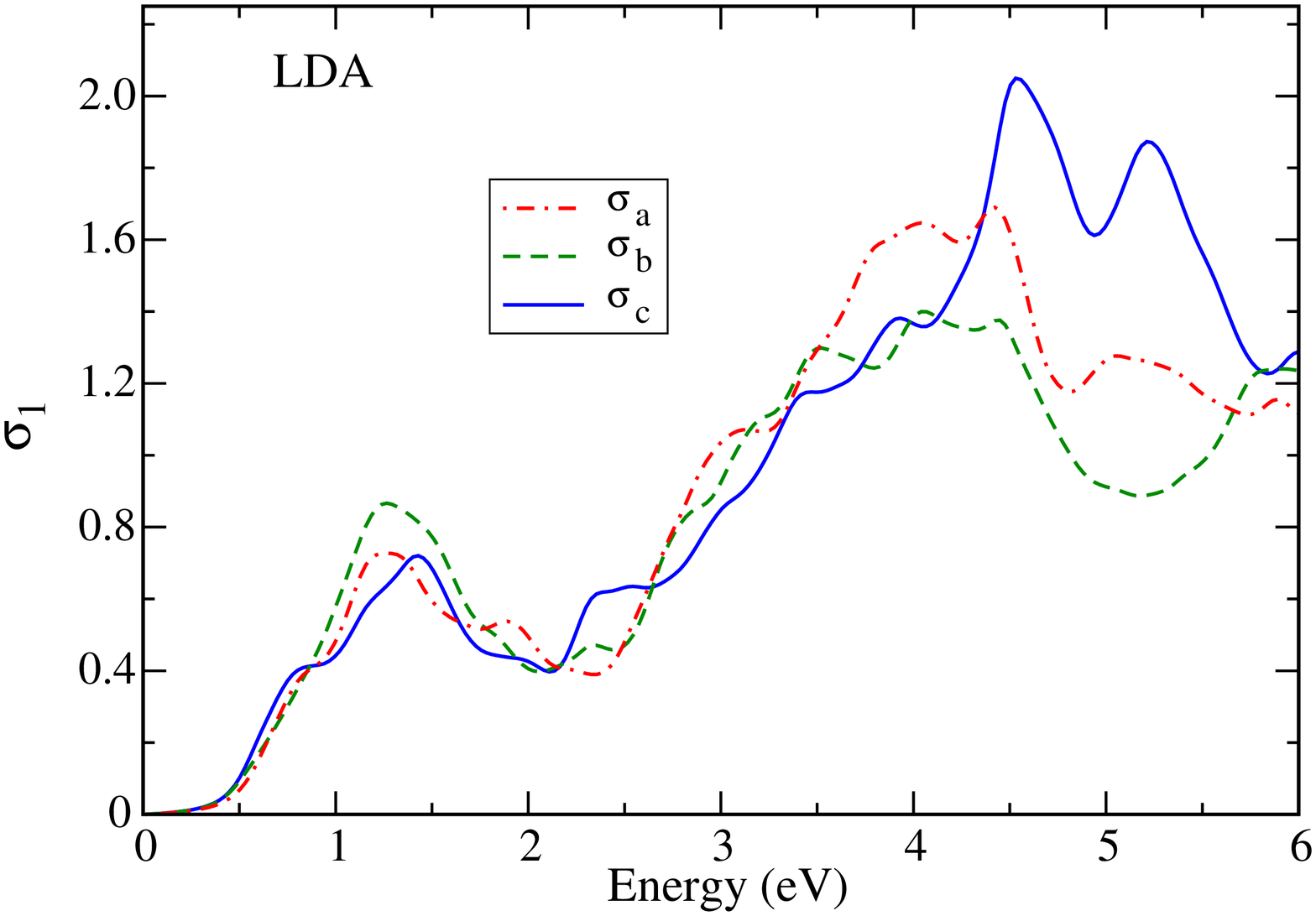}
\vspace{0.2cm} \caption{Real part of the optical conductivity within
the LSDA, as in Fig. \ref{sigma}, but with FM ordering.}
\label{sigma-fm}
\end{figure}

Figure \ref{sigma-fm} shows the calculated optical spectra for FM
ordering. As may be seen it is qualitatively similar in structure to
that of FiM ordering, but shows noticeable quantitative differences.
Most notably there are changes in the shape of the low energy Ni $d$
- $d$ peak, a suppression of the second peak and changes in the
shape of the onset of the higher energy charge transfer onset. Since
neither the FiM or the FM ordering are the ground state ordering,
these differences should be viewed as indicating the errors on our
calculation when comparing with the experimental zero field
spectrum. They also support the experimental finding of substantial
magnetochromic effects in this material. Interestingly, we find
considerable orientation dependence of the magnetochromic effects in
this material.

Optical excitations are charge excitations. Since Hubbard
correlations generally harden charge degrees of freedom leading to
spin charge separation one may expect stronger charge spin coupling,
and perhaps stronger magnetochromic effects in less correlated
materials. This is also relevant to the magnetoelectric coupling
observed in Ni$_3$V$_2$O$_8$, since strong Coulomb correlations, if
present, would reduce the spin-lattice coupling, so that while
ferroelectricity would still be induced by the incommensurate
magnetic order, it would be weaker due to reduced coupling. Based on
our results, we find that the Ni sites in Ni$_3$V$_2$O$_8$ are not
subject to strong correlations, and this underlies the substantial
magnetochromic effects seen in this material.

\subsection{Magneto-Optical Properties of Ni$_3$V$_2$O$_8$}

\begin{figure}[t]
\centerline{
\includegraphics[width=3.2in]{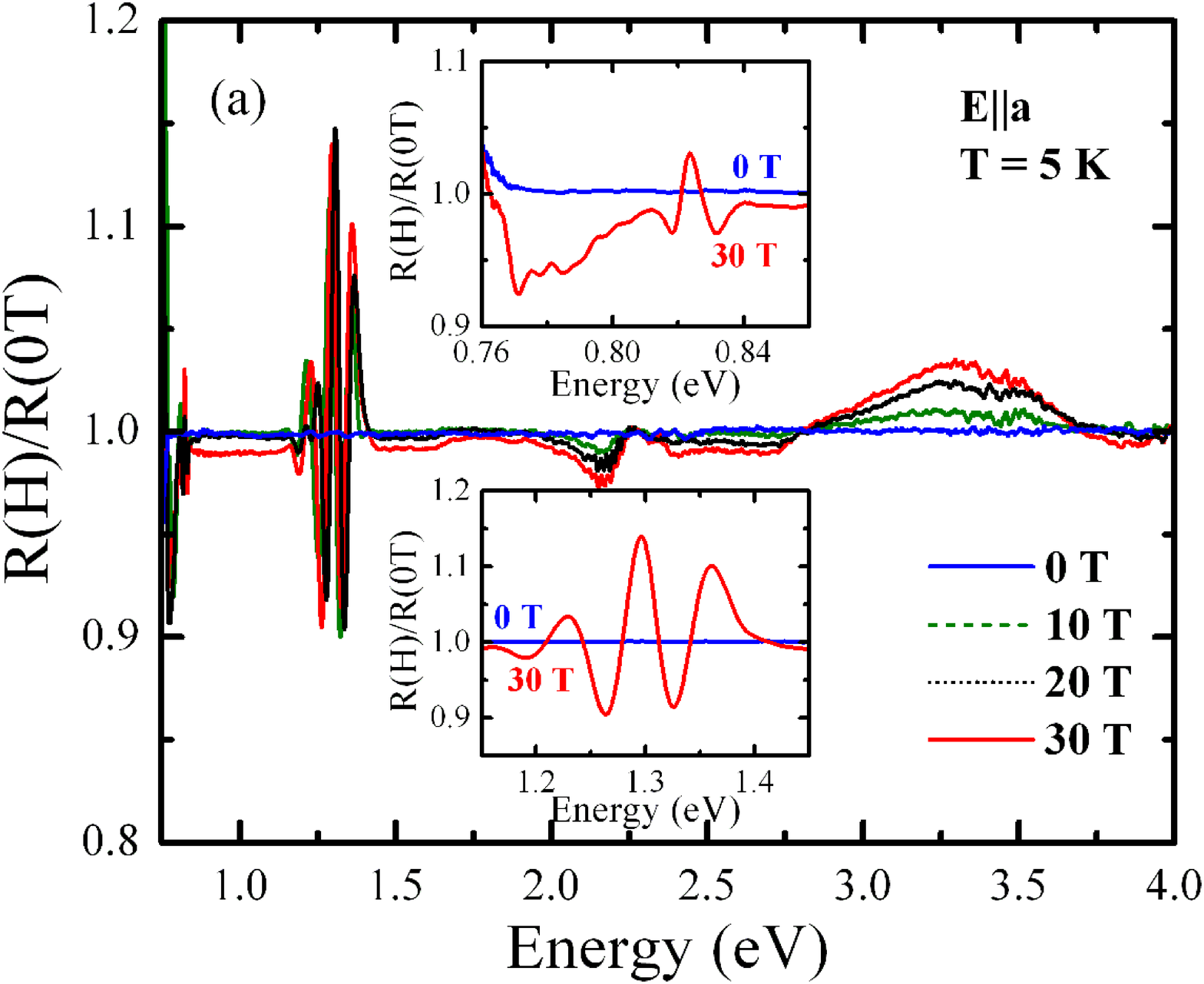}}
\vskip -1.6cm \centerline{
\includegraphics[width=3.2in]{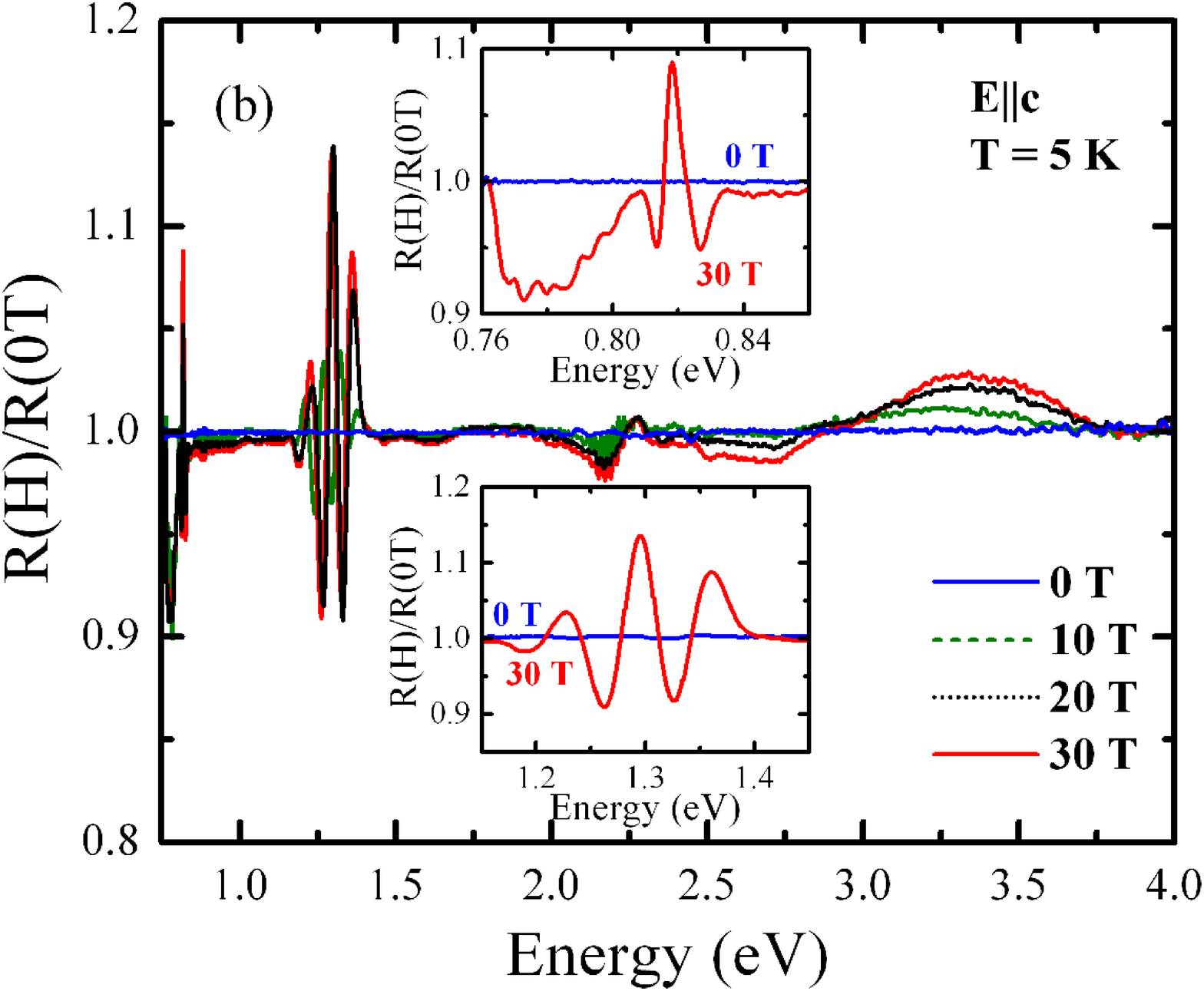}}
\vskip -0.4cm \caption{(Color online) The normalized magneto-optical
response, R($H$)/R($H$=0 T), of Ni$_3$V$_2$O$_8$ in an applied
magnetic field ($H$$\parallel$$b$) from 0 to 30 T at 5 K (a) for
light polarized along the $a$ direction. (b) for light polarized
along the $c$ direction. The insets show close-up views of the
magneto-optical response near the Ni (cross-tie) and Ni (spine) $d$
to $d$ on-site excitations. Data were taken in 2 T steps, but only
representative curves are displayed for clarity.}\label{ReflecRatio}
\end{figure}

Figure \ref{ReflecRatio}(a) and (b) shows the reflectance ratio of
Ni$_3$V$_2$O$_8$, R($H$)/R($H$=0 T), at 5 K. Rich field-induced
changes in reflectance are observed over the full energy range in
both polarizations. The most significant modifications are near 0.75
and 1.35 eV, providing a preview of the physical origin of the
field-induced effects. Based on the aforementioned positions of the
magneto-optical features, we attribute these changes to
field-induced modifications of Ni (cross-tie) $d$ to $d$ and Ni
(spine) $d$ to $d$ on-site excitations. At 30 T, the reflectance
deviates from unity  by $\pm$15\% near 1.35 eV. On the other hand,
reflectance changes in the visible range (centered at $\sim$2.2,
2.6, and 3.3 eV) are much broader and more modest in size (up to
$\sim$4\% for 30 T at 5 K), similar in magnitude to ``magnetochromic
effects" reported on other complex
materials.\cite{rai06,schnack06,woodward05,choi04} These features
can be attributed to field-induced changes in O 2$p$ to Ni 3$d$ and
O 2$p$ to V 3$d$ charge transfer excitations.

\begin{figure}[h]
\centerline{
\includegraphics[width=3.2in]{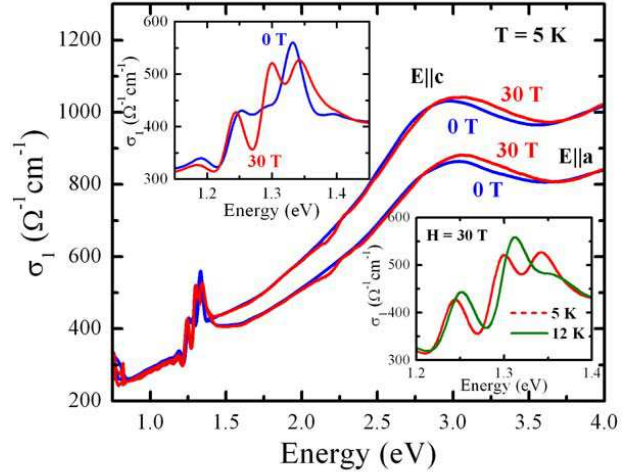}}
\vskip -0.4cm \caption{(Color online) Polarized optical conductivity
of Ni$_3$V$_2$O$_8$ at 5 K for $H$ = 0 and 30 T ($H$$\parallel$$b$).
$E$$\parallel$$a$ and $E$$\parallel$$c$ polarizations are indicated
by dashed and solid lines, respectively. The upper inset shows a
close-up view of the energy region near the Ni (spine) $d$ to $d$
on-site excitation for $H$ = 0 and 30 T. The lower inset shows a
close-up view of the Ni (spine) $d$ to $d$ on-site excitation for
$T$ = 12 and 5 K at 30 T.}\label{OpConducH30T}
\end{figure}

\begin{figure}[t]
\centerline{
\includegraphics[width=3.2in]{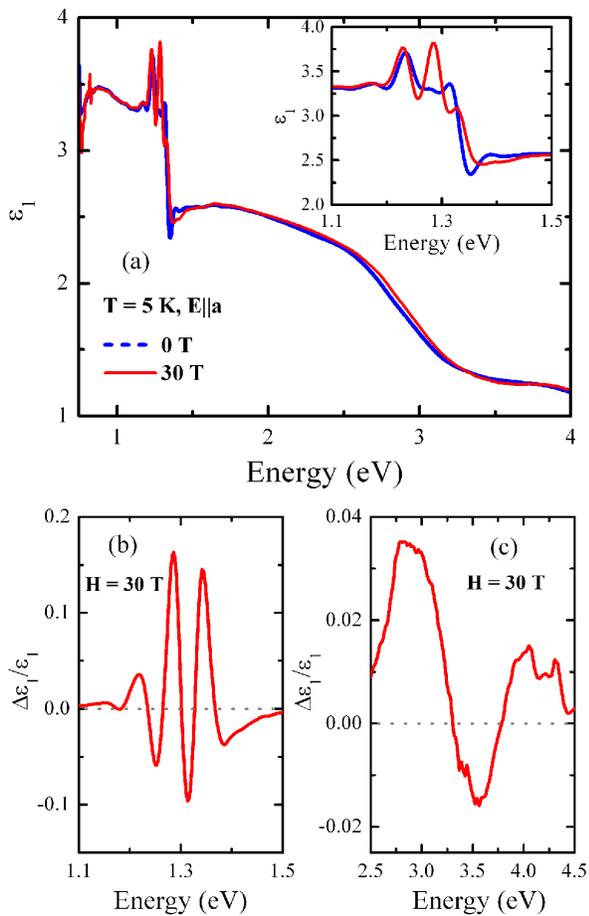}}
\vskip -0.2cm \caption{ (Color online) (a) Dielectric constant of
Ni$_3$V$_2$O$_8$ at 5 K for light polarized along the $a$ direction
for $H$ = 0 and 30 T ($H$$\parallel$$b$). The inset shows a close-up
view of the energy region near the Ni (spine) $d$ to $d$ on-site
excitation. Close-up views of the high-energy dielectric contrast,
$\Delta\epsilon_1/\epsilon_1$ = $[{\epsilon_1}(E,H) -
{\epsilon_1}(E,0)]/{\epsilon_1}(E,0)$, near (b) the Ni (spine) $d$
to $d$ on-site excitation and (c) the charge transfer
excitations.}\label{Dielec5K}
\end{figure}

It is desirable to correlate field-induced reflectance ratio changes
with the dispersive and lossy response of the material. Figure
\ref{OpConducH30T} displays the polarized optical conductivity of
Ni$_3$V$_2$O$_8$ at 5 K at 0 and 30 T ($H$$\parallel$$b$), extracted
from a combination of absolute reflectance measurements, the
reflectance ratio data of Fig. \ref{ReflecRatio}(a) and (b), and a
Kramers-Kronig analysis and calculation of the optical constants.
From the optical conductivity, we can immediately confirm that the
aforementioned magneto-optical effects correspond to field-induced
modifications of the Ni (cross-tie) $d$ to $d$, Ni (spine) $d$ to
$d$ on-site excitations, and O 2$p$ to Ni 3$d$ and O 2$p$ to V 3$d$
charge transfer excitations. The insets of Fig. \ref{OpConducH30T}
show close-up views of the $\sim$1.35 eV Ni on-site excitation
splittings that point toward a field-induced local distortion of the
NiO$_6$ octahedra.\cite{note1} As mentioned previously, these Ni $d$
to $d$ excitations are allowed because oxygen hybridization modifies
the matrix elements. The charge transfer excitation centered at
$\sim$3.0 eV also broadens and blueshifts with applied field, and a
small ``notch" develops at $\sim$ 2.2 eV. The reflectance ratio
trends discussed above also capture important dispersive effects.
Figure \ref{Dielec5K}(a) shows the high energy magneto-dielectric
response of Ni$_3$V$_2$O$_8$ at 5 K for $E$$\parallel$$a$. Although
dielectric contrast ($\Delta\epsilon_1/\epsilon_1$) is subtle in the
majority of the visible region (Fig. \ref{Dielec5K}(c)), the applied
field changes the dielectric constant significantly near the Ni $d$
to $d$ on-site excitations (inset, Fig. \ref{Dielec5K}(a), and Fig.
\ref{Dielec5K}(b)). At some
energies, dielectric contrast is on the order of 15\%
(Fig. \ref{Dielec5K}(b)). These results demonstrate that
magneto-dielectric effects are not confined to kHz frequencies and
that the response may be tunable. Both aspects
may be useful for
device applications (for instance, magnetically controlled ferroelectric memory). This remarkable interplay between the magnetic
field and optical constants of Ni$_3$V$_2$O$_8$ seems to be
facilitated by the
 fact that this material is in the intermediate
coupling regime, as discussed below.

Comparison of the reflectance ratio data in the upper insets of Fig.
\ref{ReflecRatio}(a) and (b) provides  insight into possible
magnetoeleastic coupling in Ni$_3$V$_2$O$_8$. The structure centered
at $\sim$0.83 eV is large in the $c$ (cross-tie) direction, whereas
it is much smaller along the $a$ (spine) direction. It grows quickly
with applied field. The splitting of this structure ($\sim$12 meV
(97 cm$^{-1}$)) corresponds to a vibrational energy scale.
Candidates for possible coupling might therefore include the
displacement of Ni (cross-tie) atoms or a low-frequency O-Ni-O
bending mode. In the presence of a soft lattice, magnetoelastic
coupling may lead to a local distortion of the NiO$_6$ octahedra.
Note that the PM phase of a frustrated magnetic system often
displays significant spin correlations.\cite{flores06} The PM phase
in Ni$_3$V$_2$O$_8$ may be similar, consistent with the presence of
magnetoelastic coupling. Direct measurements of high field
vibrational properties of Ni$_3$V$_2$O$_8$ are clearly desirable. We
note that spin-lattice coupling  leads to structural distortions in
the frustrated magnetic system ZnCr$_2$O$_4$.\cite{sushkov05,lee02}
Magnetoelastic coupling also  plays an important role in
geometrically frustrated magnetic systems  such as HoMnO$_3$ and
RMn$_2$O$_5$ (R = Tb, Ho, Dy).
\cite{cruz05,lorenz04,hur04,flores06,ratclif05,blake05}

\begin{figure}[t]
\centerline{
\includegraphics[width=3.2in]{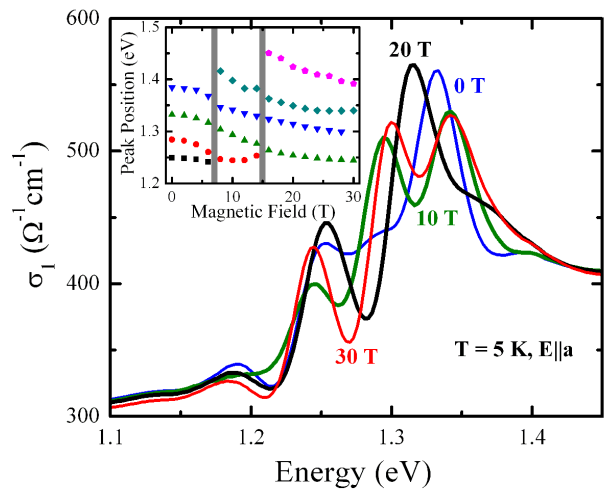}}
\vskip -0.2cm \caption{(Color online) Polarized optical conductivity
of Ni$_3$V$_2$O$_8$ at 5 K for $H$ = 0, 10, 20, and 30 T
($H$$\parallel$$b$). The inset shows the peak position of fitted
oscillators as a function of applied magnetic field. The shaded
regions represent transition fields, from which the extended $H$-$T$
phase diagram in Fig. \ref{PhaseDia}(b) was
generated.}\label{SpinePeakEa5K}
\end{figure}

Spin-charge coupling in Ni$_3$V$_2$O$_8$ can be further investigated
by quantifying the subtle changes in the Ni (spine) $d$ to $d$
on-site excitation with applied field. Figure \ref{SpinePeakEa5K}
shows  a close-up view of polarized optical conductivity of
Ni$_3$V$_2$O$_8$ at 5 K for $H$ = 0, 10, 20, and 30 T
($H$$\parallel$$b$). The effect of applied field  is non-monotonic
and correlates with changes in magnetic order. Peak fits, using four
model oscillators, were used to elucidate these trends (inset, Fig.
\ref{SpinePeakEa5K}). The dependence of peak positions on the
applied field clearly demonstrates an interplay between electronic
and magnetic properties. For instance, at 5 K, the data shows
discontinuities near 7 and 15 T, indicated by the grey shaded
regions in the inset. These transition regions move to higher fields
with increasing temperature. We note that the low field phase
boundaries determined in this manner are in good overall agreement
with the previously reported thermodynamic data. This optical
properties work also allows us to follow the C/LTI phase boundary to
higher fields, where it crosses over both the HTI and  PM phases.
Finally, these studies also identify  a new ``High Field" (HF)
region, nestled between C and the cascade of LTI, HTI, and PM phases
at higher temperature (Fig. \ref{PhaseDia}(b)). Reexamination of
previous specific heat measurements on Ni$_3$V$_2$O$_8$ are
consistent with the presence of an additional weak transition for
$H$ = 8 T, as shown by the arrow in (Fig. \ref{PhaseDia}(a)). A
similar transition was also observed at $H$ = 7 T.

As discussed in the Introduction, Ni$_3$V$_2$O$_8$ has a complex
$H$-$T$ phase diagram with several low field phases arising from the
complex spin structures.\cite{lawes05,lawes04} With the new
transition fields extracted from the magneto-optical data at 5, 8,
and 12 K (solid circles) and specific heat data (solid triangles),
we extended the $H$-$T$ phase diagram of Ref.[\onlinecite{lawes04}]
for $H$$\parallel$$b$, as shown in Fig. \ref{PhaseDia}(b). We used
the combined optical and specific heat data to establish the
boundaries for the newly identified high field phase. It has been
predicted\cite{lawes04} that the phase boundary for the LTI phase
should be quadratic in $H$, so that ($T_c(H)$-$T_c(0)$) $\sim$
$H^2$. Rather than this quadratic dependence, we find that this
phase boundary can be fit by ($T_c(H)$-$T_c(0)$) $\sim$ $H^{2.5}$.
This fit is shown by the lower solid line in Fig. \ref{PhaseDia}(b).
The origins for this discrepancy are unclear at this time.  The
upper solid line in Fig. \ref{PhaseDia}(b) is a guide to the eye for
the high field (HF) phase boundary. To the best of our
knowledge, there has been no discussion of any possible functional
form that might apply to this boundary. Combined magnetization and
specific heat measurements in high magnetic fields will be required
to precisely determine the magnetic field dependence of both phase
boundary lines, which in turn will provide mechanistic information
on the underlying magnetic transitions. Addition of the new data
points from optical and specific heat studies also demonstrates the
existence of a  new HF phase that overlays the other low field
regions.  Neutron experiments are needed to assess magnetic order in
the HF phase.

\begin{figure}[h]
\centerline{
\includegraphics[width=3.2in]{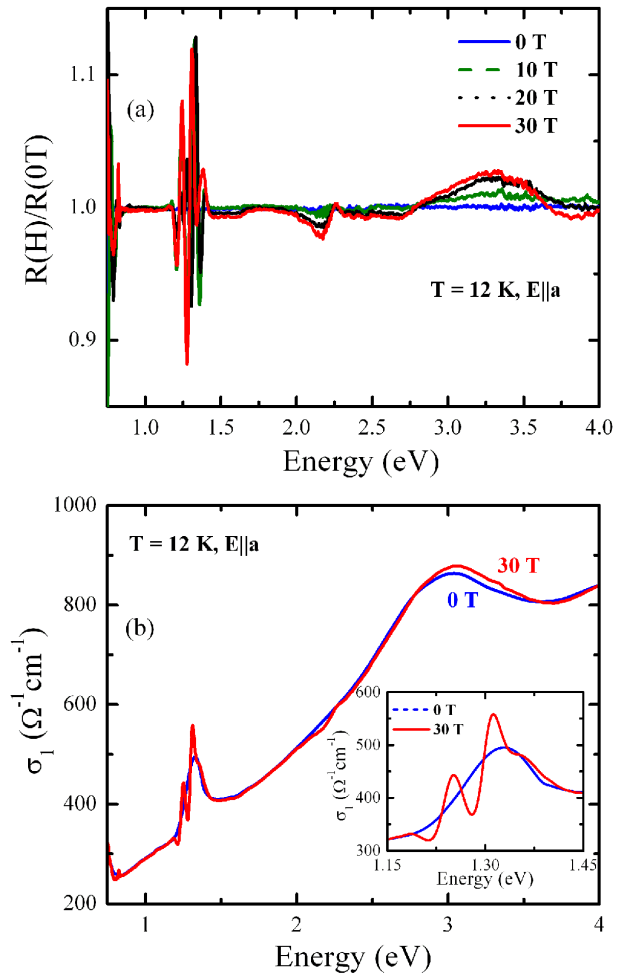}}
\vskip -0.2cm \caption{ (Color online) (a) The normalized
magneto-optical response, R($H$)/R($H$=0 T), of Ni$_3$V$_2$O$_8$ in
an applied magnetic field ($H$$\parallel$$b$) from 0 to 30 T at 12 K
for light polarized along the $a$ direction. (b) Polarized optical
conductivity of Ni$_3$V$_2$O$_8$ at 12 K for $H$ = 0 and 30 T
($H$$\parallel$$b$). Inset shows a close-up view of the
magneto-optical response near the  Ni (spine) $d$ to $d$ on-site
excitation.}\label{OpConduc12K}
\end{figure}

In addition to the 5 K (LTI phase) data discussed in detail here, we
also collected full data sets in the HTI and PM phases of
Ni$_3$V$_2$O$_8$. Strong magneto-optical effects were observed at
both 8 and 12 K. As shown in Fig. \ref{OpConduc12K}, the
magneto-optical effects in the PM phase are qualitatively similar to
those of the LTI and HTI (not shown) phases, although the overall
size of the field-induced modification is smaller. This result
demonstrates that an applied field induces a local structural
distortion around the Ni sites in all of the low temperature phases.
In otherwords, the field-induced distortion is not unique to the LTI
(ferroelectric) phase with spontaneous polarization. To support this
claim, we refer the reader to the inset of Fig.
\ref{OpConduc12K}(b)), where the zero-field $\sim$1.35 eV excitation
is featureless. An applied field  splits this structure into several
distinct components, demonstrating the field-induced distortion of
the Ni (spine) crystal field environment even in the PM phase.

\section{Conclusion}

We investigated the optical properties, electronic structure, and
energy dependent magneto-optical response of Ni$_3$V$_2$O$_8$ to
elucidate the electronic structure and to study the phase diagram.
The spectra exhibit features centered at $\sim$0.75 and $\sim$1.35
eV that we assign as Ni (cross-tie and spine) $d$ to $d$ on-site
excitations in the minority spin channel.  O 2$p$ to Ni 3$d$ and O
2$p$ to V 3$d$ charge transfer excitations appear at higher energy.
Extensive analysis of splitting patterns of the Ni (spine) $d$ to
$d$ excitation in the PM, HTI, and LTI phases demonstrates that the
local Ni environment is sensitive to magnetic order even at zero
field. A splitting of Ni (cross-tie) $d$ to $d$ on-site excitation
between 50 and 100 K also points toward a weak local structural
distortion that precedes the cascade of low temperature magnetic
phases. Although both Ni$_3$V$_2$O$_8$ and the prototypical Mott
insulator NiO are based on Ni$^{2+}$ ions octahedrally coordinated
by O with similar bond lengths, the basic electronic structures of
these two materials are very different. NiO has a large band gap and
local moments that derive from strong Coulomb interactions and
separation of spin and charge degrees of freedom, while, in
contrast, we found Ni$_3$V$_2$O$_8$ to be an intermediate gap, local
moment band insulator. This electronic structure is particularly
favorable for magneto-dielectric coupling, because the material is
not subject to the spin charge separation of the strongly correlated
large gap Mott insulator, while at the same time, remaining a
magnetic insulator independent of the particular spin order and
temperature. The remarkable interplay of magnetic field and optical
properties in Ni$_3$V$_2$O$_8$ reveals an additional high field
phase and an unexpected electronic structure, which we associate
with the strong magnetodielectric couplings in this material. We
discovered several prominent magneto-optical effects that derive
from changes in crystal field environment around Ni (spine and
cross-tie) centers due to a field-induced modification of local
structure of NiO$_6$. The magnetoelastic mechanism, responsible for
the field-induced distortion of the NiO$_6$ building block unit, is
active in the paramagnetic phase as well.

\section{ACKNOWLEDGMENTS}

Work at the University of Tennessee is supported by the Materials
Science Division, Basic Energy Sciences, U.S. Department of Energy
(DE-FG02-01ER45885). Research at ORNL is sponsored by the Division
of Materials Sciences and Engineering, Office of Basic Energy
Sciences, U.S. Department of Energy, under contract
DE-AC05-00OR22725 with Oak Ridge National Laboratory, managed and
operated by UT-Battelle, LLC. Work at UC Davis is supported by DOE
grant DE-FG03-01ER45876. A portion of this research was performed at
the NHMFL, which is supported by NSF Cooperation Agreement
DMR-0084173 and by the State of Florida. Work at Princeton
University is supported by NSF through the MRSEC program (NSF MRSEC
grant DMR-9809483). We appreciate useful discussions with W. E.
Pickett, and thank Q. Huang and M. Woodward for their assistance in
preparing Fig. \ref{struct}.

\vfill\eject


\begin{thebibliography}{100}
\bibitem{u-note}
Here we used an effective $U-J$ = 5 eV, $J$ = 0, though because of
the fully spin polarized nature of the Ni the results would not
depend on the choice of $J$ provided $U-J$ is held constant.
\bibitem{note1}
Although magneto-optical measurements below 0.75 eV are limited by
the energy range (0.75 - 4.1 eV) of our experimental set-up, we
suspect that  splitting of Ni(cross-tie) $d$ to $d$ excitation may
be similar to the splitting of Ni (spine) $d$ to $d$ excitation.
\end{thebibliography}
\end{document}